\documentclass[%
 aip,
 amsmath,amssymb,
reprint, 
]{revtex4-1}

\usepackage{graphicx}
\usepackage{dcolumn}
\usepackage{bm}

\usepackage[utf8]{inputenc}
\usepackage[T1]{fontenc}
\usepackage{mathptmx}
\usepackage{etoolbox}
\usepackage{color}
\makeatletter
\def\@email#1#2{%
 \endgroup
 \patchcmd{\titleblock@produce}
  {\frontmatter@RRAPformat}
  {\frontmatter@RRAPformat{\produce@RRAP{*#1\href{mailto:#2}{#2}}}\frontmatter@RRAPformat}
  {}{}
}%
\makeatother
\begin{document}

\preprint{AIP/123-QED}

\title{Noncollinear phase-matching of high harmonic generation in solids}
\author{Pavel Peterka}
\author{František Trojánek}
\author{Petr Malý}
\author{Martin Kozák}
\email{m.kozak@matfyz.cuni.cz}
\affiliation{Department of Chemical Physics and Optics, Faculty of Mathematics and Physics, Charles University, Ke Karlovu 3, 12116 Prague 2, Czech Republic}

\date{\today}

\begin{abstract}
We propose and experimentally demonstrate a scheme allowing to reach noncollinear phase-matching of high harmonic generation in solids, which may potentially lead to an enhancement of the generation efficiency. The principle is based on high-order frequency mixing of two light waves with identical frequencies but different directions of wavevectors. In this process, $N$-th harmonic frequency is produced by frequency mixing of $N$+1 photons from a wave with high amplitude of electric field and a single photon from a wave with low field amplitude, which are propagating noncollinearly in an optically isotropic media. We experimentally verify the feasibility of this scheme by demonstrating phase-matched generation of third and fifth harmonic frequency in sapphire.
\end{abstract}

\maketitle

\section{Introduction}
High harmonic generation (HHG) is a highly nonlinear optical process in which the interaction of high intensity optical field with matter leads to frequency up-conversion. When the interaction takes place in the nonperturbative regime, the spectra generated by interband polarization and intraband currents contain a rich information about the illuminated material. HHG in solids has been used to investigate various properties of materials such as the band structure\cite{vampa2015all,lanin2017mapping,suthar2022}, Berry curvature\cite{luu2018measurement}, topological surface states\cite{bai2021high}, the dynamics of photoinduced phase transitions\cite{bionta2021tracking}, coherent phonon dynamics\cite{Neufeld2022} or it was use to perform microscopic imaging\cite{zayko2021,korobenko2023situ}. The HHG process requires high intensity of the driving laser and it typically has low conversion efficiency, which can be improved in solids by fabricating special nanostructures. Studies have demonstrated that locally enhanced fields in nanostructures with subwavelength dimensions can significantly improve HHG efficiency. Various principles have been applied to reach the local field enhancement such as the use of geometrical conical shapes \cite{sivis2017tailored,franz2019all}, plasmonic resonances \cite{vampa2017plasmon,imasaka2018antenna,an2021efficient,jalil2022controlling}, grating structures \cite{liu2020beating} and resonant dielectric metasurfaces \cite{liu2018enhanced,shcherbakov2021generation,zograf2022high,peterka2023high}.

The less explored way how to increase HHG efficiency in solids is based on the phase matching improvement. To increase the efficiency of HHG, the microscopic sources in different depths of the sample have to constructively interfere in the far field. However, in an optically isotropic media, the values of the refractive index at the fundamental and harmonic frequencies are typically different due to dispersion leading to a nonzero phase mismatch for collinear HHG of $\Delta \textbf{k} = N\textbf{k}^{\omega}-\textbf{k}^{N\omega}$, where $N$ is the harmonic order and $\textbf{k}^{\omega}$ and $\textbf{k}^{N\omega}$ are the wavevectors of the fundamental and harmonic waves (Figure \ref{setup}a left). In second-order nonlinear optical processes, phase matching is typically achieved in optical anisotropic media, where the dispersion curves for waves with different directions of linear polarization corresponding to ordinary and extraordinary rays can be adjusted to fulfil the phase-matching condition. In the simplest case of second harmonic generation by a single optical wave, type I phase-matching (oo-e or ee-o) leads to perpendicular polarizations of the waves at the fundamental and second harmonic frequencies requiring nonzero off-diagonal elements of the second-order nonlinear susceptibility. In the case of higher-order nonlinear susceptibilities, the values of off-diagonal elements decrease significantly with the increasing nonlinearity order. Moreover, when generating higher harmonics, the phase-mismatch becomes larger due to an increasing difference between the frequencies of the driving and generated waves preventing to fulfil the phase-matching condition even in anisotropic media.  

An alternative technique which can be applied in optically isotropic media is quasi-phase matching. Its principle is based on spatial modulation of the efficiency of harmonic generation in the material. For second-order nonlinear processes, the periodic switching of the direction of effective nonlinear coefficient is typically realized by using a periodically poled ferroelectric crystal. Here the second-order nonlinear susceptibility is modulated due to the periodically inverted orientation of crystal axis \cite{armstrong1962interactions,gordon1993diffusion,hickstein2017high}. Periodically poled lithium niobate waveguide was used to generate quasi-phase matched harmonics up to 13th order corresponding to a wavelength of 315 nm\cite{hickstein2017high}.

In addition to medium modulation, quasi-phase matching can also be achieved by modulating the generation efficiency along the beam propagation by interference with another light wave, whose amplitude of electric field can be much smaller than the amplitude of the driving wave. Series of counterpropagating pulses lead to periodic modulation of the intensity, which can control HHG in gases \cite{peatross1995intensity,peatross1997selective,voronov2001control}, second harmonic generation \cite{lytle2016influence} or HHG in solids \cite{korobenko2019high}.  
The schemes which involve two noncollinear beams offer another degree of freedom allowing to tune the phase mismatch. By changing the angle between the two beams, the phase matching condition can be fulfilled for a specific combination of the generated harmonic frequency and the propagation direction of the generated wave leading to angular separation of beams at harmonic frequencies \cite{bertrand2011ultrahigh,heyl2014macroscopic,jiang2021multi}.

Besides the two waves at the same frequency, the HHG in solids has been controlled by two color scheme with the fundamental and its second harmonic frequency. It allows to break the inversion symmetry and hence both the odd and even harmonic frequencies are generated in centrosymmetric materials \cite{luu2018observing,tzur2022selection}. On top, the polarization state of HHG can be controlled by chiral pump beams allowing to generate circularly polarized harmonics with various selection rules\cite{heinrich2021chiral, rana2022generation}. An enhancement of HHG efficiency has also been theoretically studied in various solid materials \cite{song2020enhanced, navarrete2020two, avetissian2022efficient}.

Here we propose and experimentally demonstrate a scheme for noncollinear phase-matching of HHG in solids. The scheme is based on difference frequency generation of $N$+1 photons of strong beam and 1 photon of weak beam to generate $N$-th harmonic at specific angle (Figure \ref{setup}a right). The feasibility of the method is experimentally verified by phase matching the third and fifth harmonic generation processes in sapphire.  

\section{Theory}

The proposed phenomenon can be described in the framework of classical nonlinear optics. The phase-matching condition for the wave at the third harmonic frequency $3\omega$ can be derived from the wave equation in nonlinear media.\cite{boyd2008nonlinear} 
\begin{equation}
\Delta \tilde{E}_3 -\frac{\varepsilon^{(1)}(3\omega ) }{c^{2}}\frac{\partial^{2} \tilde{E}_3}{\partial t^{2}}=\frac{1}{\varepsilon_0 c^{2} }\frac{\partial^{2} \tilde{P}_{NL}}{\partial t^{2}},
\end{equation}
where $\varepsilon^{(1)}$ is linear part of dielectric function, $\tilde{E}_3$ is complex electric field amplitude of the wave at third harmonic frequency and $\tilde{P}_\text{NL}$ is nonlinear polarization. Scalar approximation of the fields is used because we assume that all the waves have linear polarization in the direction perpendicular to the plane of incidence.
The electric field of the two noncollinear driving waves at frequency $\omega$ and the generated wave at frequency 3$\omega$ is assumed to have the form of plane waves:
\begin{equation}
\begin{array}{l}
\tilde{E}_1(\mathbf{r},t)=\tilde{E}_{10} e^{i(\omega t-\mathbf{k_1\cdot r})},\\
\tilde{E}_2(\mathbf{r},t)=\tilde{E}_{20} e^{i(\omega t-\mathbf{k_2\cdot r})},\\
\tilde{E}_3(\mathbf{r},t)=\tilde{E}_{30}(\mathbf{r})e^{i(3\omega t-\mathbf{k_3\cdot r})},
\end{array}
\end{equation}
where $\tilde{E}_{10}$, $\tilde{E}_{20}$ and $\tilde{E}_{30}(\mathbf{r})$ are the envelopes of the three waves and $\mathbf{k_1}$, $\mathbf{k_2}$ and $\mathbf{k_3}$ are their wave vectors. The length of the wave vectors in the material is ${k}_1=k_2=2\pi n_1/\lambda$ and ${k}_3=3\pi n_3/\lambda$, where $n_1$ is the refractive index of the material at the fundamental wavelength $\lambda$ and $n_3$ is the refractive index at the wavelength $\lambda/3$ corresponding to the third harmonic frequency. Because the generation efficiency is low, we assume negligible depletion of the incident waves ($\tilde{E}_{10}$ and $\tilde{E}_{20}$ are constants). We consider the generation of the third harmonic frequency by a 5th order frequency mixing process which involves four photons $(N+1)$ from the strong wave and 1 photon from the weak wave because this is the only term of the total nonlinear polarization, for which the generation can be phase-matched. In the case of the third harmonic generation by third order nonlinear polarization, the argument of the complex exponential contained in the coupled wave equation (see the derivation of Eq. (4)) cannot be zero for any combination of wavevectors $\pm a \mathbf{k_1} \pm b \mathbf{k_2} \pm \mathbf{k_3}$, where $a$ and $b$ are positive integer numbers fulfilling the relation $a+b=3$. The only term which contributes to the nonlinear polarization for phase-matched generation of the third harmonic frequency thus can be written using the fifth-order nonlinear optical susceptibility $\chi^{(5)}$ as:
\begin{equation}
\tilde{P}^{(5)}_{NL}(3\omega=4\omega-\omega)=5 \epsilon_0 \chi^{(5)} \tilde{E}_1^{4}\tilde{E}^*_2,
\end{equation}
where asterisk symbol denotes the complex conjugate. Nonlinear polarization obtained using Eq. (3) is used as a source in Eq. (1). Further we apply the slowly-varying envelope approximation, we assume that $\left| \tilde{E}_{10}\right| \gg \left| \tilde{E}_{20} \right|$ and that the generated wave at frequency 3$\omega$ propagates along $z$ direction. Solution of Eq. (1) leads to the coupled wave equation: 
\begin{equation}
\frac{\partial \tilde{E}_{30}}{\partial z} =\frac{45\tilde{E}_{10}^{4}\tilde{E}^*_{20}\omega^{2}\chi^{(5)} }{2k_3ic^{2}}e^{-i(4\mathbf{k_1 \cdot r}-\mathbf{k_2\cdot r}-{k_3 z})}.
\end{equation}
After integrating Eq. (4) from the $z=0$ to $z=L$ we obtain the amplitude $\tilde{E}_{30}(L)$. The resulting intensity of the wave at third harmonic frequency $I_3 =\frac{cn_3\varepsilon_0}{2}\left|\tilde{E}_3\right|^2$ can be written as:
\begin{equation}
I_3=\frac{(45)^2 4I_{1}^{4}I_{2}\omega^{2}\chi^{(5)2}}{\varepsilon_0^4c^6n_3n_1^5}L^2 \text{sinc}^2\left(\frac{\Delta \textbf{k}L}{2}\right).
\end{equation}
When the phase-matching condition $4\mathbf{k_1\cdot r}-\mathbf{k_2\cdot r}-{k_3 z}=0$ is fulfilled, the intensity of the third harmonic wave $I_{3}$ increases quadratically with $L$. In the plane wave approximation the $L$ represents thickness of the crystal. In experiments with focused optical beams with finite pulse duration, the generated harmonic intensity depends also on the spatio-temporal overlap of the pulsed beams. We can generalize the phase-matching condition for generation of $N$-th harmonic frequency, which reads:
\begin{equation}
(N+1)\mathbf{k_1\cdot r}-\mathbf{k_2\cdot r}-{k^{N\omega}_3 z}=0,  
\end{equation}

where $k^{N\omega}_3$ is the length of the wavevector of the wave at $N$-th harmonic frequency.

The presented phase-matching scheme can also be understood from a different point of view. The interference of the two beams at the fundamental frequency leads to modulation of nonlinear polarization in the medium (see Fig. \ref{setup}c), which in turn causes periodic modulation of the effective nonlinearity. When the period of the modulation along the propagation direction of the generated wave $\Lambda=2\pi / (k_{1,z}-k_{2,z})$ is equal to twice the coherence length for the generation of a specific harmonic frequency $L_\text{coh}=2/\Delta k$, the harmonic generation is periodically switched on and off leading to a coherent build up of the generated wave. From this perspective, the working principle is similar to quasi-phase matching in periodically poled crystals \cite{armstrong1962interactions,gordon1993diffusion,hickstein2017high}, which can be described using the theory presented in \cite{boyd2008nonlinear} giving formally equivalent results to the description using coupled wave equations. The main difference to the standard quasi-phase matching scheme is that there is an angle between the constant intensity fronts of the interference pattern of the two waves at the fundamental frequency and the wavefronts of the generated wave $\tilde{E}_3$ (see Fig. \ref{setup}c). After integration over the finite transverse width of the generated beam (infinite in our case as we assume the plane wave approximation), the step-like dependence of the electric field amplitude $\tilde{E}_3$ on the propagation distance expected for quasi-phase matching (see chapter 2.4 in \cite{boyd2008nonlinear}) is modified to the linear dependence obtained from Eq. (4) when assuming perfect phase-matching.

\section{Experimental}

\begin{figure*}[ht!]
\centering\includegraphics[width=11cm]{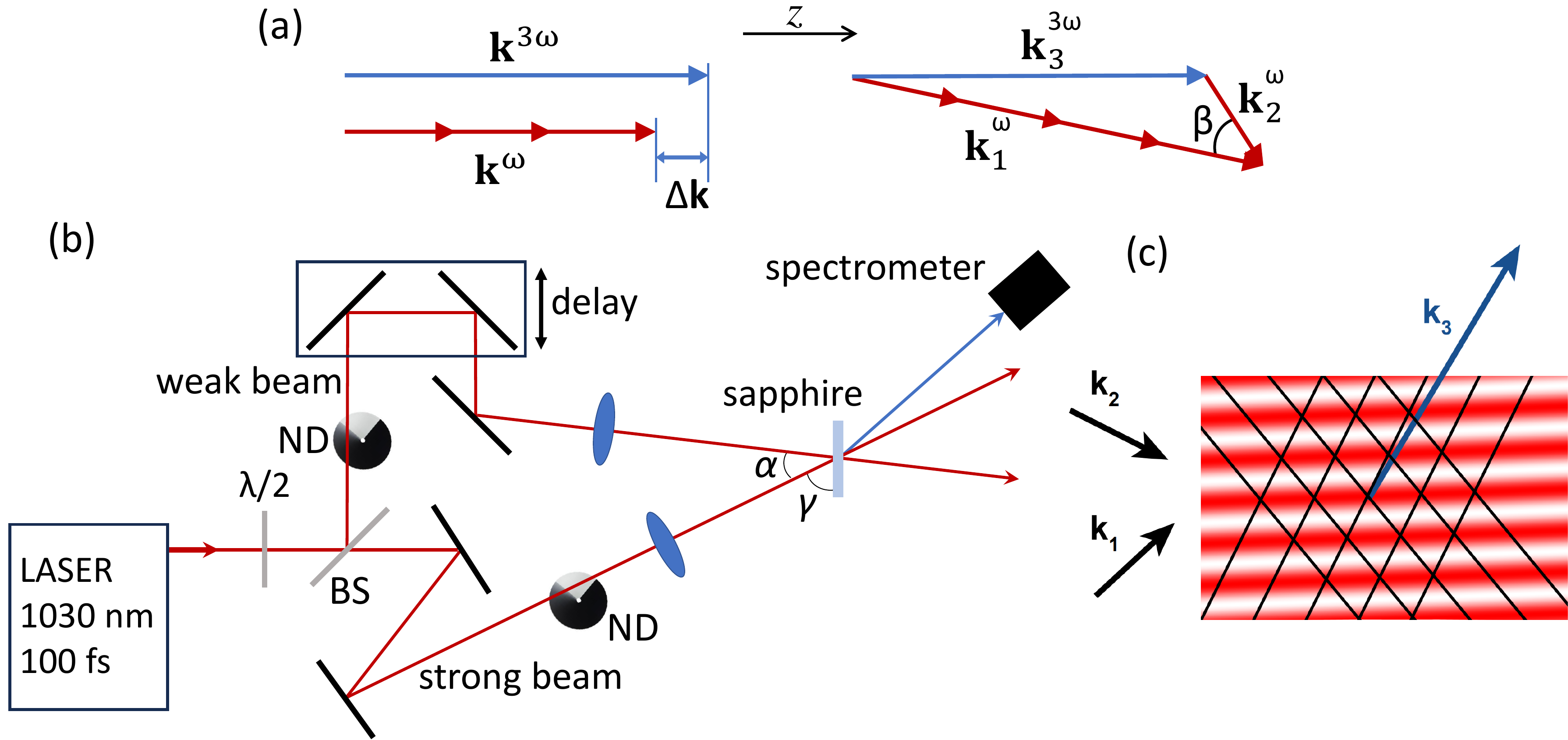}
\caption{(a) Left: Phase mismatch during collinear third harmonic generation in a dispersive
medium. Right: Phase-matched generation of third harmonic frequency using two
noncollinear beams by utilizing fifth-order nonlinear optical susceptibility $\chi^{(5)}(3\omega=4\omega-\omega)$. (b) Layout of the experimental setup. Laser beam is split to two arms, which are
focused to the sapphire sample. The angle between the two beams is $\alpha$ and the angle
between the strong beam and sample surface is $\gamma$. $\lambda/2$, half-wave plate; BS, beam splitter; ND, neutral density filter. (c) Sketch of the interference of the two waves. The intensity modulation leads to phase-matched third harmonic frequency at particular angle.}
\label{setup}
\end{figure*}

The feasibility of the proposed method is experimentally tested by phase-matched generation of the third and the fifth harmonic frequencies in a sapphire crystal with the thickness of 1 mm. We use femtosecond laser pulses with the central wavelength of 1030 nm and  the duration (FWHM) of 170 fs at the repetition rate of 25 kHz. The pulses are compressed using spectral broadening induced by self-phase modulation in a 1 mm thick fused silica plate with antireflective coating placed inside a multi-pass cell (MPC)\cite{schulte2016nonlinear, goncharov2023few}. The cell consists of two concave spherical mirrors with focal lengths of 15 cm at a distance of 59 cm. The beam passes 20 times through the fused silica window, whose position with respect to the focus of the beam in the cell is optimized to reach the best shape of the spectrum. Self-phase modulation due to the nonlinear propagation in fused silica leads to pulses with positively chirped central part. By using negatively chirped input pulses we compensate this effect and obtain pulses with duration of 100 fs (as measured by frequency resolved optical gating \cite{stibenz2005interferometric}), which are used in the experiments (see the layout of the experimental setup in Fig.~\ref{setup}b).  

The direction of linear polarization of both beams is adjusted by a half-wave plate to be perpendicular to the plane of incidence. This configuration leads to the strongest interference between the electric field components of the two waves inside the sapphire crystal. The beam is split into two parts by a beamsplitter with split ratio 20:80 and the power levels in both arms are controlled by metallic neutral density filters. The time delay between the two pulses is controlled by change of the optical path of the weak beam using delay line. The high- and low-intensity beams are focused to the crystal using lenses with focal lengths of $f_1$=15 cm and $f_2$=20 cm to focal beam radii (1/$e^2$) of 48 {\textmu}m and 63 {\textmu}m, respectively. The sample is mounted on a 2D translation stage in a holder which allows its rotation around vertical axis. The generated harmonic beam is collimated and focused to a grating spectrometer (Andor Shamrock 163) equipped with cooled CCD camera (Andor iDUS 420).

The angle between the two beams inside the sapphire crystal is set in order to satisfy the phase matching condition. From the right image in Fig. \ref{setup}a we get the formula: 
\begin{equation}
\cos\beta =\frac{(N+1)^{2}k_1^{2}+k_2^{2}-k_3^{2}}{2(N+1)k_1k_2}.  
\label{cos}
\end{equation}
 
 In the experiment we keep the angle $\alpha$ between the two generating beams constant. The angle $\beta$ between the waves inside the sample is controlled by rotating the sample surface with respect to the strong beam. When we define the angle $\gamma$ as the angle between the wave vector of the strong beam $\mathbf{k_1}$ and the sapphire surface (see Fig. \ref{setup}b), the angle $\beta$ inside the sample can be expressed as:
 
 \begin{equation}
\beta = \sin^{-1}\frac{\cos \gamma}{n}-\sin^{-1}\frac{\cos(\alpha+\gamma) }{n},
\label{beta}
\end{equation}

For our analysis of phase-matching we use the dispersion of refractive index of sapphire from Ref.\cite{weber1986handbook}. We note that the crystal is cut such that the optical axis is perpendicular to the surface and thus all the beams feel the ordinary refractive index of sapphire. When considering symmetric case, in which $2\gamma+\alpha=\pi$, phase-matching of third harmonic generation with the fundamental wavelength of 1030 nm requires $\alpha=35^{\circ}$. We adjust this angle in the experiment and keep it constant.  

\section{Results and discussion}

\begin{figure*}[ht!]
\centering\includegraphics[width=14.5cm]{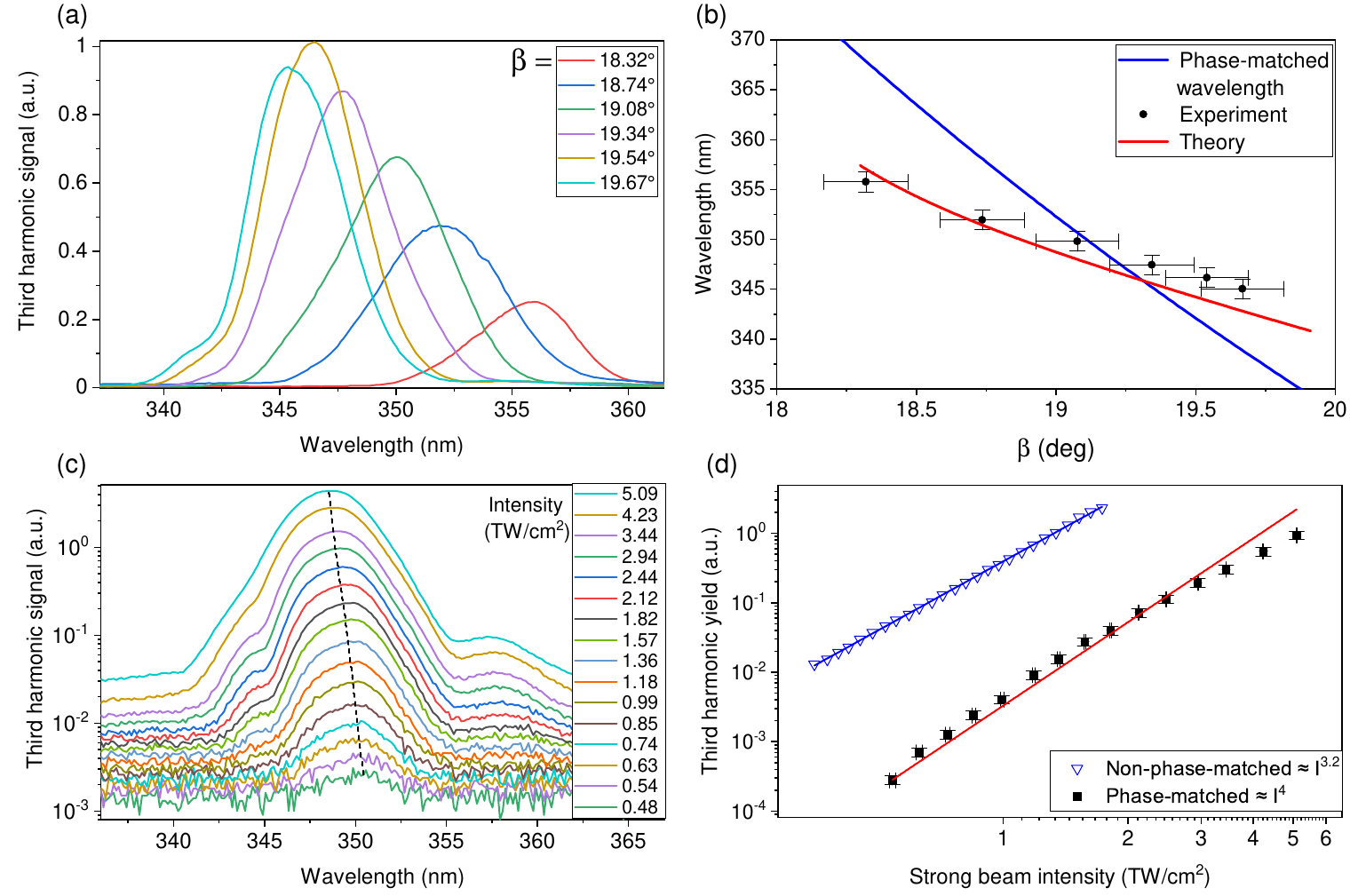}
\caption{(a) Measured spectra of phase-matched third harmonic generation in sapphire. The angle between the two beams outside of the sample is fixed to $\alpha=35^{\circ}$ and the angle $\gamma$ between the sample surface and the wave vector of the strong beam is changed by rotating the sample, which leads to the change of angle $\beta$. (b) Phase-matched wavelength (blue). Theoretically (red curve) and experimentally (black dots) obtained dependence of the central wavelength of the third harmonic frequency signal on the angle $\beta$ between the two beams inside the sapphire crystal. (c) The spectra measured for different intensities of the strong beam for fixed angle $\beta=19.67^{\circ}$. (d) The yield of third harmonic generation as a function of the peak intensity of the strong beam inside the sapphire in phase-matched configuration (rectangles) fitted by function $f(I)=aI^4$ compared to the yield from single strong beam at perpendicular incidence (triangles). Peak intensity of the weak beam in the phase-matched case is 0.23 TW/cm$^2$.}
\label{3spectra}
\end{figure*}

Third harmonic generation spectra measured with different values of the angle $\beta $ between the beams inside of the sample are shown in Fig. \ref{3spectra}a. Peak intensities of the fundamental pulses are 4.2 TW/cm$^2$ and 0.23 TW/cm$^2$, respectively. The observed shift of the spectral maxima of the third harmonic signal is caused by changing the angle $\beta$ between the two fundamental waves inside the sample leading to a shift of the phase-matched wavelength, which can be calculated by applying the phase-matching condition given by Eq. (6).     

The central wavelengths of the generated third harmonic frequency obtained by fitting the spectra shown in Fig. \ref{3spectra}a by a Gaussian function are shown in Fig. \ref{3spectra}b (squares) as a function of the angle $\beta$. To find the corresponding theoretical dependence of the wavelength corresponding to maximum spectral power density of the generated third harmonic frequency we first calculate the phase-matched wavelength using Eqs. (\ref{cos}) and (\ref{beta}) (blue curve in Fig. \ref{3spectra}b). To take into account the finite phase-matching bandwidth and the finite spectral width of the pulses at the fundamental frequency, the third harmonic spectrum is assumed to have Gaussian envelope and the phase matching is approximated by a sinc squared function coming from Eq. (5) centered at the phase-matched wavelength. For each angle $\beta$ we calculate the generated third harmonic spectrum as a product of the Gaussian envelope and the sinc squared function. The wavelength of the peak of the calculated spectrum is shown as the red curve in Fig. \ref{3spectra}b. The agreement between the measured and theoretical results is influenced by the experimental conditions (complex time and spatial overlap of the fundamental pulses in the sample and precision of determination of the angle $\beta$).    
\begin{figure*}[ht!]
\centering\includegraphics[width=12cm]{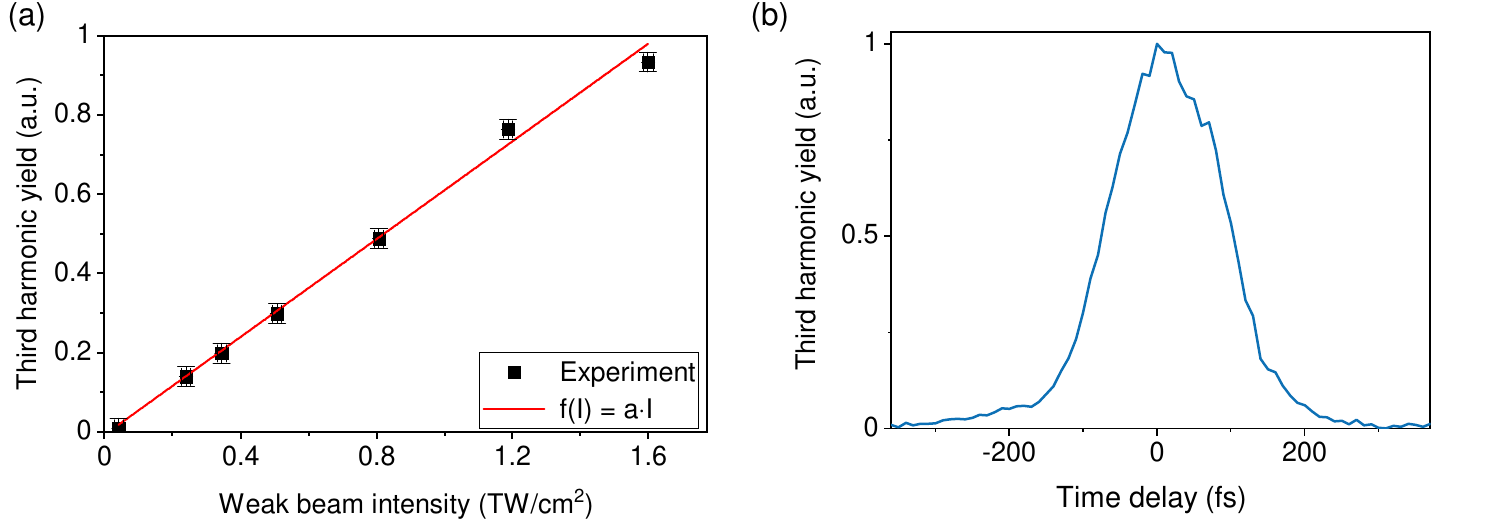}
\caption{(a) The yield of third harmonic generation as a function of the peak intensity of the weak beam fitted by linear function. Peak intensity of the strong beam is 5.1 TW/cm$^2$. (b) Dependence of the yield of third harmonic generation on the time delay between the two generating pulses.}
\label{3b}
\end{figure*}  

For a fixed peak intensity of the weak beam of 0.23 TW/cm$^2$ and increasing intensity of the strong beam, the THG spectrum shifts to shorter wavelengths (see Fig. \ref{3spectra}c). This is caused by the nonlinear refractive index $\Delta n\approx \chi^{(3)}\left | E_{01} \right|^2$, which leads to an increase of the length of the wave vector of the fundamental beam with increasing peak intensity. To satisfy the phase-matching condition (see Fig. \ref{setup}a right), the wave vector of THG needs to increase in magnitude, which corresponds to the shift to shorter wavelengths.

To further verify the origin of the signal at the third harmonic frequency in the fifth-order nonlinear process we characterize the dependence of the yield of third harmonic generation on the peak intensities of both beams at the fundamental frequency. The light intensities inside the sapphire sample are calculated from the measured peak intensities of the incident pulses beams assuming the Fresnel losses at the surface. The THG yield corresponds to the integral of the measured THG spectral power density shown in Fig. \ref{3spectra}c. Its dependence on the intensity of the strong beam for a fixed peak intensity of the weak beam of 0.23 TW/cm$^2$ is shown in Fig. \ref{3spectra}d. The dependence is shown on log-log scale along with the function $f(I)=aI^4$ corresponding to the process in which four photons of the strong beam are absorbed to create one photon at the third harmonic frequency. The dependence of THG yield on the peak intensity of the weak beam with fixed peak intensity of the strong beam of 5.1 TW/cm$^2$ is linear (see Fig. \ref{3b}a) implying the involvement of a single photon from the weak beam in the generation process. We verified that the intensity dependence of the standard non-phase-matched THG in the direction of the strong beam is $I_\text{THG}\approx I^{3.2}$ (Fig. \ref{3spectra}d). Another way to verify the origin of the third harmonic signal in the high order frequency mixing is by controlling the time delay between the weak and strong pulses at the fundamental frequency. The measured third harmonic signal is present only when the pulses overlap in time in the sample. The dependence of THG yield on the time delay (Fig. \ref{3b}b) shows Gaussian-like shape with FWHM=154 fs roughly corresponding to the cross-correlation of the 100 fs pulses in the sapphire sample, which is influenced by the spatio-temporal overlap of the optical fields.
\begin{figure*}[ht!]
\centering\includegraphics[width=14cm]{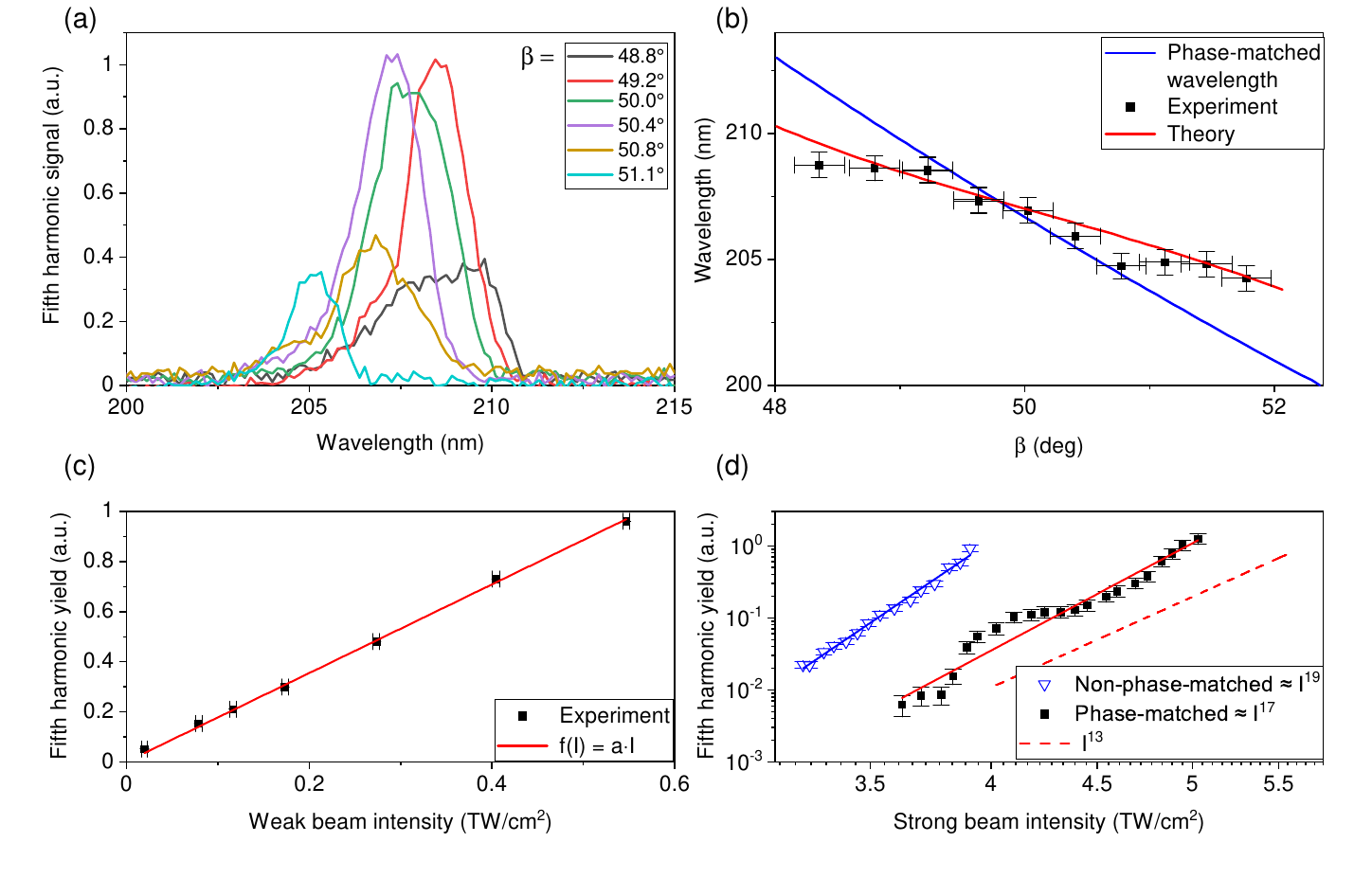}
\caption{(a) Measured spectra of phase-matched fifth harmonic generation in sapphire for different angles $\beta$ between the two beams inside the sapphire crystal. (b) Phase-matched wavelength (blue). Theoretically (red curve) and experimentally (black dots) obtained dependence of the central wavelength of the fifth harmonic frequency signal on the angle $\beta$. (c) The yield of fifth harmonic generation as a function of the peak intensity of the weak beam fitted by linear function. Peak intensity of the strong beam is 5 TW/cm$^2$. (d) The yield of fifth harmonic generation as a function of the peak intensity of the strong beam inside sapphire in phase-matched configuration (rectangles) compared to the yield from single strong beam at perpendicular incidence (triangles). Intensity of the weak beam in the phase-matched case is 0.55 TW/cm$^2$. The dependencies are fitted with function $f(I)=aI^{b}$ with results $b_{NPM}=19$ and $b_{PM}=17$. The dashed line with a slope of $I^{13}$ is shown for comparison.}
\label{5spectra}
\end{figure*}  

Similar set of experiments are repeated for the fifth harmonic generation, which in our case involves frequency mixing of six photons of the high intensity beam and one photon of the low intensity beam. For the fifth harmonic frequency at 206 nm, Eqs. (\ref{cos}) and (\ref{beta}) give the minimal angle between the two fundamental beams $\alpha=97^{\circ}$. Because of the high refractive index of sapphire at this wavelength of $n$=1.9, the angle of incidence of the generated fifth harmonic beam at the output interface is close to the critical angle of total reflection $\alpha_\text{cr}=32^{\circ}$ leading to high Fresnel losses. To decrease the Fresnel reflection of the fifth harmonics when propagating out of the crystal we use a larger value of the angle $\alpha=108^{\circ}$. The measured fifth harmonic spectra with peak intensities 5 TW/cm$^2$ and 0.55 TW/cm$^2$, respectively, are shown in Fig. \ref{5spectra}a as a function of the angle $\beta$. The spectral shift of the generated fifth harmonics with the angle indicates its origin in the phase-matched interaction similar to the case of third harmonic generation. 
 
The dependence of the measured wavelength corresponding to the maximum of the generated 5th harmonic spectra on the angle $\beta$ between the waves inside the sapphire crystal is shown as black squares in Fig. \ref{5spectra}b along with the phase-matched wavelength (blue curve) and the calculation, which takes into account the finite spectral widths of the fundamental pulse and the phase-matching condition (red curve).

The dependence of the yield of the fifth harmonic generation on the peak intensity of the low intensity beam is confirmed to be linear (Fig. \ref{5spectra}c). When we change the peak intensity of the strong beam, in the perturbative regime of nonlinear optics we expect the power law $\approx I^6$. At high intensity corresponding to the nonperturbative interaction regime (Keldysh parameter $\gamma<1$ )\cite{Keldysh1965}, the exponent in the power law typically decreases. However, the measured dependence of the fifth harmonic yield on the intensity of the strong beam in sapphire scales with a much higher exponent of $\sim I^{17}$ (see Fig. \ref{5spectra}d). 

\begin{figure}[ht!]
\centering\includegraphics[width=2.5cm]{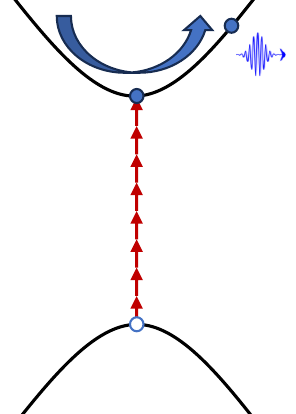}
\caption{Schematic picture of the multiphoton ionization and fifth harmonic generation. 8-photon absorption creates electron-hole pair, which is accelerated by the laser field and generates high harmonic photons.}
\label{band}
\end{figure}

To improve our understanding of this phenomenon we performed two additional experiments. First, we generated the 5th harmonic frequency directly by a single beam, which resulted in similar nonlinear order in the intensity dependence. Second, we measured the pump and probe experiment to determine the excited electron population as a function of the pump intensity (see Supplementary Figure 1). The resulting dependence on the pump intensity was again similar ($I^{18}$). The perturbative fifth harmonic generation should depend on the fifth power of the driving pulse intensity. Because we observe the dependence of the third harmonics $I^{3.2}$ in the same region of peak intensities, propagation effects of the fundamental beam are excluded as those would also influence the effective nonlinearity in the case of third harmonic generation. Therefore we conclude that the high nonlinear order has a microscopic origin. The multiphoton generation of an electron-hole pair to the conduction and valence bands in sapphire\cite{temnov2006multiphoton} requires 8 photons with photon energy of 1.2 eV because the band gap of sapphire is 8.8 eV \cite{french1990electronic}). Our data thus suggest that the real carriers play role in the observed fifth harmonic generation process. The only possibility how the effective nonlinearity can be increased is that the fifth harmonics in this regime is mainly generated by the intraband current due to electrons excited via multiphoton absorption of 8 photons to the conduction band (see the sketch in Fig. \ref{band}). The subsequent fifth order nonlinearity is the consequence of anharmonic motion of the excited electrons in the band, which would add another factor of 5 to the exponent in the intensity dependence. We note that with such strongly nonlinear intensity dependence, the difference between $I^{13}$ and $I^{17}$-$I^{19}$ is rather small, which is demonstrated by adding the curve for $I^{13}$ to the Fig. \ref{5spectra}d. We note that high harmonic spectra generated in sapphire by few-cycle laser pulses \cite{Kim2017} did not show such high order nonlinear dependence of the driving pulse intensity.

The strong beam intensity dependence of fifth harmonics shows non-monotonic behavior as shown in Fig. \ref{5spectra}d). Similar behaviour has been shown in GaAs \cite{xia2021high}, but it can not explain our result, because we do not observe the non-monotonic intensity dependence for single beam fifth harmonic generation at normal incidence. The non-monotonic behaviour of the intensity dependence in the case of phase-matched harmonic generation is thus most probably related to the phase matching and to nonlinear modification of Fresnel losses due to nonlinear refractive index of the material.

Regarding the generation efficiency of the phase-matched scheme, Figures \ref{3spectra}d and \ref{5spectra}d show the comparison of the harmonic generation yield with the standard single beam interaction. The phase-matched generation yield is about 1-2 orders of magnitude lower in the case of our experiments. There are two reasons causing that we do not observe an enhancement of the HHG yield by phase-matching. The first reason is that the phase-matched generation is based on nonlinear optical susceptibility of higher order than the non-phase-matched generation, which makes the energy transfer to the nonlinear polarization and the generated wave weaker. The second factor limiting the phase-matched generation efficiency in our experiments is the short distance of only few tens of micrometers over which the generation occurs. The generation distance can be in future significantly extended by optimizing the spatio-temporal overlap of the pulses for example by using pulse front tilt. When we assume the prolongation of the phase-matched distance by a factor of 100-times to approx. 1 mm, the enhancement of the phase-matched generation efficiency based on the results of our theoretical calculations (Eq. (5)) is by a factor of $10^4$ due to the dependence on the length of the generating media $I_N \approx L^2$ (this dependence is general also for the higher nonlinear orders in the approximation of negligible depletion of the strong wave). The noncollinear phase-matching scheme demonstrated here would be beneficial in particular in the nonperturbative regime, where the HHG spectra typically contain a plateau with similar spectral power densities of several subsequent harmonic orders. When both $N$-th and $N$+2-th harmonic orders are in the plateau region, the noncollinear phase-matching scheme would lead to an enhancement of the $N$-th order harmonic generation.

The presented phase-matching scheme can be generalized to the nondegenerate case by utilizing two beams at different frequencies. When the phase-matching cannot be reached by identical frequencies due to excessively large difference between the wavevectors of the waves at the fundamental and high harmonic frequencies, one can mix the fundamental frequency with one of its low-order harmonics to fulfill the phase-matching condition. 

In our experiments we use a plan-parallel crystal and the beams are incident on one of the planar parallel surfaces. When the phase-matching angle between the beams becomes large, it would be beneficial to use another geometry with prismatic shape of the crystal, in which an arbitrary value of the angle $\beta$ can be reached. The Fresnel losses at surfaces can be minimized by carefully selecting the geometry of the prismatic sample to achieve perpendicular incidence of the beams. Also, the crystal surface can be coated with antireflective coating for specific fundamental and high harmonic frequencies. 

Besides increasing the generation efficiency, the presented noncollinear phase-matching scheme brings also opportunities for controlling the polarization, propagation direction or optical angular momentum of the selected harmonic order as it has been demonstrated previously\cite{hickstein2015non,kong2017controlling,heinrich2021chiral}. 

\section{Conclusion}
In summary, we have introduced and experimentally investigated a technique allowing to reach phase-matching of high harmonic generation in solids using two non-collinear beams. The phase-matching condition has been derived from the nonlinear wave equation for the process of third harmonic generation and can be straightforwardly extended to $N$-th harmonic generation. The phase-matched interaction was confirmed by measuring the spectra generated in sapphire as a function of the angle between the two generating beams, their relative time delay and their peak intensities. We note that the maximum photon energy of the generated harmonic frequency is limited by the presence of electronic resonances in the material, which should be transparent for both the fundamental and the generated frequencies. By using wide band gap crystals, the photon energy can reach vacuum ultraviolet spectral region (for example 14 eV in LiF). Nonperturbative high harmonic spectra typically contain a plateau, where the intensity of several subsequent harmonic orders have similar spectral power densities. In this regime, the yield of generation can be significantly enhanced by the proposed phase-matching scheme  even when using the process which requires $N+2$-order nonlinearity for $N$-th harmonic generation, providing that both $N\omega$ and $(N+2)\omega$ are in the plateau region.       

\section*{Supplementary material}
See the supplementary material for the results of pump probe experiment.

\begin{acknowledgments}
The authors would like to acknowledge the support by Charles University (Nos. UNCE/SCI/010, SVV2020-260590, PRIMUS/19/SCI/05, GA UK 1190120) and the Czech Science Foundation (No. 23-06369S), co-funded by the European Union (ERC, eWaveShaper, No. 101039339). Views and opinions expressed are, however, those of the author(s) only and do not necessarily reflect those of the European Union or the European Research Council. Neither the European Union nor the granting authority can be held responsible for them. This work was supported by TERAFIT project No. CZ.02.01.01/00/22\_008/0004594 funded by OP JAK, call Excellent Research. 
\end{acknowledgments}

\section*{Author Declarations}
\subsection*{Conflict of Interest}
The authors have no conflicts to disclose.
\subsection*{Author Contributions}
Pavel Peterka: Formal analysis; Investigation; Writing – original draft. František Trojánek: Software (lead), Writing – review \& editing (equal). Petr Malý: Writing – review \& editing (equal). Martin Kozák: Conceptualization (lead); Methodology; Supervision (lead); Writing – review \& editing (equal).

\section*{Data Availability Statement}

The data that support the findings of
this study are available from the
corresponding author upon reasonable
request.

\section*{References}
\bibliography{aipsamp}

\end{document}